\documentclass[journal=nat,manuscript=perspective]{achemso}

\usepackage{graphicx} 
\usepackage[version=3]{mhchem} 

\usepackage{array}
\usepackage{caption}
\usepackage{amsmath}
\usepackage{graphicx}
\usepackage{longtable}
\usepackage[dvipsnames, table]{xcolor}
\usepackage[T1]{fontenc}
\usepackage[colorlinks=true, allcolors=blue]{hyperref}

\newcolumntype{P}[1]{>{\centering\arraybackslash}p{#1}}

\title{Computational Methods to Investigate Intrinsically Disordered Proteins and their Complexes}
\author{Zi Hao Liu*}
\affiliation{Molecular Medicine Program, Hospital for Sick Children, Toronto, Ontario M5G 0A4, Canada}
\altaffiliation{Department of Biochemistry, University of Toronto, Toronto, Ontario M5S 1A8, Canada}
\author{Maria Tsanai*}
\affiliation{Kenneth S. Pitzer Center for Theoretical Chemistry and Department of Chemistry, University of California, Berkeley,  Berkeley, California 94720, USA}
\author{Oufan Zhang}
\affiliation{Kenneth S. Pitzer Center for Theoretical Chemistry and Department of Chemistry, University of California, Berkeley,  Berkeley, California 94720, USA}
\author{Julie Forman-Kay}
\affiliation{Molecular Medicine Program, Hospital for Sick Children, Toronto, Ontario M5G 0A4, Canada}
\altaffiliation{Department of Biochemistry, University of Toronto, Toronto, Ontario M5S 1A8, Canada}
\author{Teresa Head-Gordon}
\affiliation{Kenneth S. Pitzer Center for Theoretical Chemistry and Department of Chemistry, University of California, Berkeley,  Berkeley, California 94720, USA}
\altaffiliation{Departments of Bioengineering and Chemical and Biomolecular Engineering, University of California, Berkeley,  Berkeley, California 94720, USA}

\email{thg@berkeley.edu}

\date{August 2024}

\begin{document}
\maketitle

\begin{abstract}
\noindent
In 1999 Wright and Dyson highlighted the fact that large sections of the proteome of all organisms are comprised of protein sequences that lack globular folded structures under physiological conditions. Since then the biophysics community has made significant strides in unraveling the intricate structural and dynamic characteristics of intrinsically disordered proteins (IDPs) and intrinsically disordered regions (IDRs). Unlike crystallographic beamlines and their role in streamlining acquisition of structures for folded proteins, an integrated experimental and computational approach aimed at IDPs/IDRs has emerged. In this Perspective we aim to provide a robust overview of current computational tools for IDPs and IDRs, and most recently their complexes and phase separated states, including statistical models, physics-based approaches, and machine learning methods that permit structural ensemble generation and validation against many solution experimental data types. 
\end{abstract}
\newpage
\section{Introduction}
Despite the widely accepted protein structure-function paradigm central to folded proteins, it is increasingly appreciated that all proteomes also encode intrinsically disordered proteins and regions (IDPs/IDRs), which do not adopt a well-defined 3D structure but instead form fluctuating and heterogeneous structural ensembles.\cite{Wright1999, Dyson2005} The so-called "Dark Proteome" is made up of IDPs/IDRs that comprise over 60\% of proteins in eukaryotes\cite{Bhowmick2016}, and this abundance together with growing experimental evidence challenge the assumptions that protein function and protein interactions require stable folded structures.\cite{Tsang2020}

Proteins with intrinsic disorder confers certain advantages over folded protein states. Plasticity of disordered protein states facilitates conformational rearrangements and extended conformations that allow them to interact simultaneously with multiple spatially separated partners, changing shape to fold upon binding or in creating dynamic complexes.\cite{Wright1999} Disordered regions in protein complexes may control the degree of motion between domains, permit overlapping binding motifs, and enable transient binding of different binding partners, facilitating roles as signal integrators and thus explaining their prevalence in signaling pathways.\cite{Forman-Kay2013} Disorder is also highly over-represented in disease-associated proteins, and IDPs have been shown to be involved in a variety of fundamental processes including transcription, translation and cell cycle regulation that when altered lead to cancer and neurological disorders\cite{Iakoucheva2002}. 

Recent evidence suggests that IDPs/IDRs are enriched in biomolecular condensates.\cite{Brangwynne2011, Lin2015} Biomolecular condensates arise from phase separation, percolation and other related transitions \cite {Mittag2022} to induce a biomacromolecule-rich phase and a dilute phase depleted of such biomacromolecules \cite{Hyman2014, Boeynaems2018}, a phenomenon well-established by polymer physics\cite{Flory1953, Brangwynne2009}. IDPs/IDRs have been suggested to promote phase separation and other related transitions due to their structural plasticity, low-complexity sequence domains, and multivalency.\cite{Wright2015} Furthermore, functional dynamic complexes of IDPs/IDRs and biomolecular condensates are known to be sensitive to post-translational modifications (PTMs).\cite{KimTae2019} Regulatory PTMs\cite{Zhang2024} often target residues within IDPs/IDRs, as they are more accessible and flexible than folded protein elements\cite{Darling2018}. Modification of IDPs/IDRs by PTMs is well known to modulate many cellular processes, including dynamic complexes \cite{Mittag2010} and the assembly/disassembly, localization, and material properties of biomolecular condensates \cite{Snead2019}.
 
\begin{figure}
  \includegraphics[width=\textwidth]{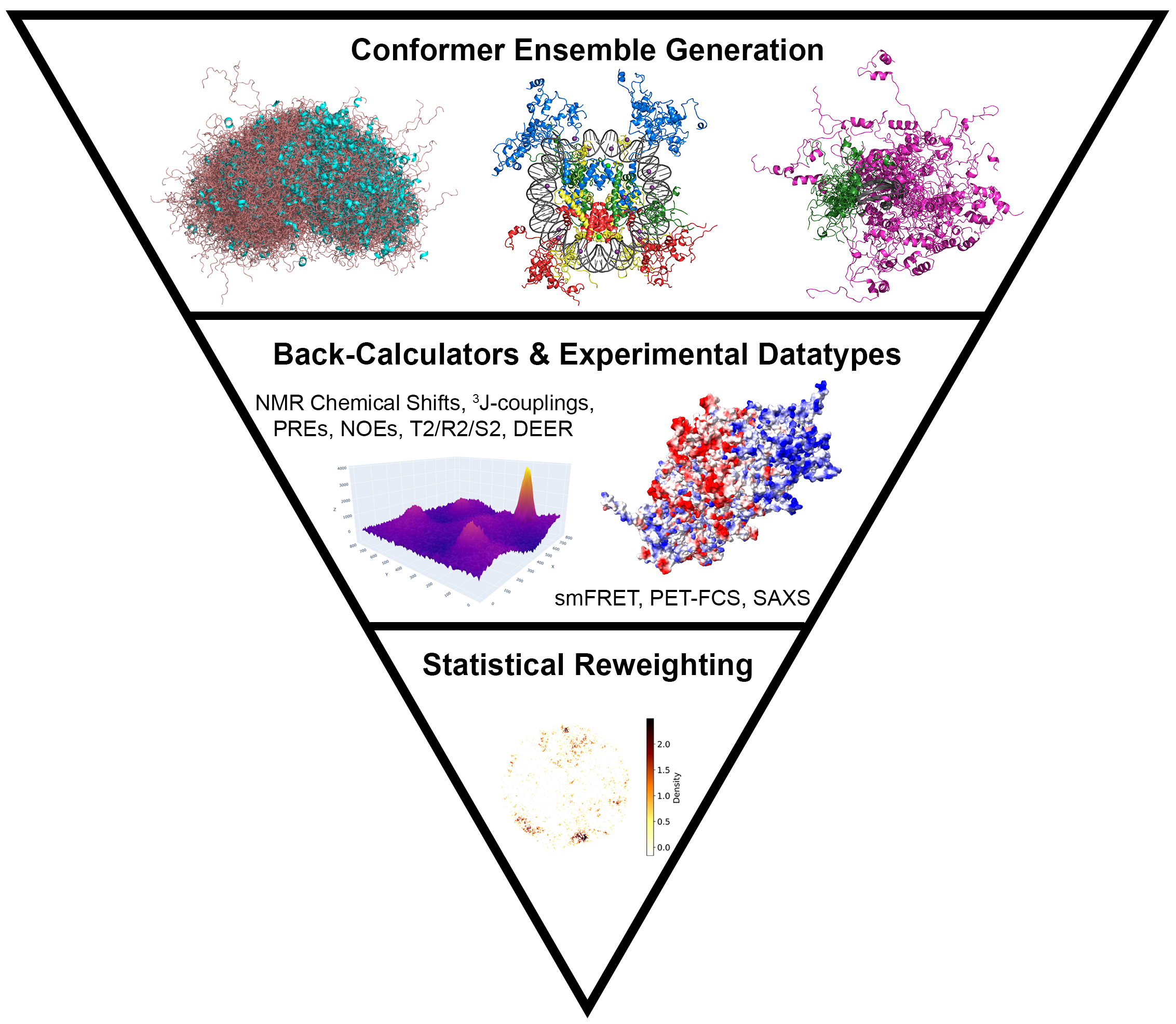}
  \caption{\textit{Schematic of the IDP/IDR conformer ensemble generation pipeline.} (top) Initial conformers are generated by using knowledge- and/or physics-based methods. Examples from IDPConformerGenerator are shown from left-to-right, ensembles of the Inhibitor-2 IDP (loop regions in salmon, helices in cyan), nucleosome core particle adapted from PDB ID 2PYO (histone H2A in yellow, H2B in red, H3 in blue, H4 in green), and 5-phosphroylated folded 4E-BP2 adapted from PDB ID 2MX4 (N-terminal IDR in green, C-terminal IDR in magenta). (middle) 3D schematic of an NMR 2D HSQC is shown along with an electrostatic surface model of the I-2 conformers. (bottom) A density heatmap is shown with N = 1500 conformations of the 5p 4E-BP2 ensemble with illustrations done using ELViM\cite{Viegas_2024}. Figure adapted with permission from Oxford University press.}
  \label{fig:IDPIDR}
\end{figure}

One of the primary goals in understanding disordered proteins and their roles in biology is to create, validate, and analyze IDP/IDR structural ensembles to imbue insight into the conformational substates that give rise to IDP/IDR function (Figure \ref{fig:IDPIDR}). Folded proteins have well-defined experimental approaches, mostly using X-ray crystallography, electron crystallography and microscopy, and recent computational approaches such as AlphaFold2\cite{Jumper2021} can determine their structure with high accuracy. IDPs/IDRs bring new challenges to both experiment and computer simulation and modeling, requiring an integrative biology approach of experimental and computational methods that work together to characterize their diverse and dynamic structural ensembles. 

In this perspective, we review the computational tools for structural ensemble creation and validation for isolated IDPs/IDRs, including those that operate by generating and evaluating structural ensembles that are consistent with the collective experimental restraints derived from Nuclear Magnetic Resonance (NMR), Small Angle X-ray Scattering (SAXS), and other available solution experimental measurements. But to fully address the biological activity of IDPs/IDRs, we must further develop computational methods to characterize dynamic complexes and biological condensates, and to account for changes due to PTMs. Here we provide a current snapshot of available computational methods, workflows, and software that are available for characterizing IDPs and IDRs and their complexes, while also identifying current gaps for future progress in their characterization.

\section{Ensemble Generation}
Generating a conformational ensemble can be categorized into knowledge-based approaches, physical models, and machine learning (ML)/artificial intelligence (AI) methods. Knowledge-based generation protocols are a popular choice as ensembles can be calculated in a few minutes on a workstation or laptop\cite{Teixeira2022}. Although computationally more expensive, all-atom (AA) and coarse-grained (CG) physical models combined with molecular dynamics (MD) simulations are also widely used, becoming increasingly more accurate when using many-body force fields, better algorithms for sampling, and rapid quarterly developments on silicon hardware as well as parallel application-specific integrated circuits (ASICs) like the Anton 3 supercomputer.\cite{Shaw2021} Finally, ML/AI tools have exploded onto the scene with improved generative models being utilized to predict structural ensembles of disordered proteins.

\subsection{Knowledge-Based Methods}
Knowledge-based ensemble generation approaches have three things in common: 1) Building protein chains using fragments of residues to retain the local information of a modeled peptide. 2) Exploiting a database of high-resolution non-redundant PDB structures as an empirical force-field. 3) Being exceptionally computationally efficient while generating free conformers without unphysical steric clashes. Since the early 2000s, there have been many software packages released for modeling disordered proteins using statistical or static methods. TraDES, introduced as FOLDTRAJ, models IDP structures \cite{Feldman2000} using a 3-residue fragment chain growth protocol, derived from a curated database of non-redundant protein structures from the RCSB PDB\cite{Berman2000}. \textit{Flexible-meccano} creates statistical ensembles by randomly sampling amino acid-specific backbone torsion ($\phi$/$\psi$) angles from high-resolution X-ray crystallographic protein structures, and attempts to validate them with experimental NMR as well as SAXS data\cite{Ozenne2012}. 

More recently, the FastFloppyTail method allows for the prediction of disordered ensembles by exploiting the Rosetta AbInitio protein prediction model \cite{Ferrie2020}, for not only isolated IDPs but also for IDRs at the termini of folded structures. The latest software platform for modeling and analyzing IDPs/IDRs, IDPConformerGenerator, can also use NMR and SAXS experimental data to bias conformer generation, but unlike the other methods allows for the generation of ensembles in different contexts such as transmembrane systems, dynamic complexes, and applications in biological condensates \cite{Teixeira2022, LiuLDRS2023} (see Figure \ref{fig:IDPIDR}). IDPConformerGenerator also includes the ability to generate side chain ensembles using the Monte Carlo Side Chain Ensemble (MCSCE) method\cite{Bhowmick2015} including with PTMs\cite{Zhang2024} for IDPs/IDRs and their complexes.

\subsection{Physics-Based Methods}
MD has been extensively used to study the conformations, dynamics, and properties of biological molecules, and can reproduce and/or interpret thermodynamic and spectroscopic data, and can also provide accurate predictions for processes inaccessible to experiment\cite{Fawzi2021, Shea2021}. Because most AA force fields used in MD were developed to represent folded proteins, they can lead to inaccuracies when studying disordered proteins and regions. Thus, intensive work has been dedicated to refining well-established "fixed charge" force fields (FFs), resulting in notable improvements. The D. E. Shaw group employed six fixed charge FFs with explicit solvent to study the structural properties of both folded and disordered proteins \cite{Robustelli2018}. By modifying the torsion parameters and the protein-water interactions, the a99SB-disp FF\cite{Robustelli2018} was shown to provide accurate secondary structure propensities for a plethora of disordered proteins, while accurately simulating folded proteins as well. However, recent studies have reported that the a99SB-disp model is too soluble for studying the condensation of some disordered proteins \cite{Arghadwip2021, Samantray2020}. Jephtah \textit{et al.} utilized different versions of the CHARMM \cite{Liu2019} and AMBER \cite{LindorffLarsen2010} FFs with different water models \cite{Piana2015, Robustelli2018} to study five peptides that exhibit a polyproline II helix (PPII) structure \cite{Jephthah2021}. Interestingly, most models managed to capture, to varying extents, the ensemble averages of the radius of gyration and the PPII secondary structure, but the models differed substantially in the under- or over-sampling of different secondary structures.

Recently, Liu and co-workers considered a new direction – the connection of FFs to configurational entropy – and how that might qualitatively change the nature of our understanding of FF development that equally well encompasses globular proteins, IDPs/IDRs, and disorder-to-order transitions.\cite{Liu2021} Using the advanced polarizable AMOEBA FF model, these many-body FFs generate the largest statistical fluctuations consistent with the radius of gyration (Rg) and universal Lindemann values (correlated with protein melting temperature) for folded states. These larger fluctuations of folded states were shown to translate to their much greater ability to simultaneously describe IDPs and IDRs such as the Hst-5 peptide, the stronger temperature dependence in the disorder-to-order transition for (AAQAA)$_3$, and for maintaining a folded core for the TSR4 domain while simultaneously exhibiting regions of disorder.\cite{Liu2021} This supports the development and use of many-body FFs to described folded proteins and to create IDP/IDR ensembles, by naturally getting the energy-entropy balance right for all biomolecular systems. 

CG models reduce the resolution of an all-atom model in order to simulate larger systems and for longer timescales, which often is necessary when considering disordered proteins and dynamic complexes and condensates. Heesink \textit{et al.} used a CG model that represents each amino-acid with a single interaction site or "bead" to investigate the structural compactness of $\alpha$-synuclein \cite{Heesink2023}. The SIRAH model represents only the protein backbone with three beads and can be applied to both folded and disordered proteins \cite{Klein2021}. Joseph \textit{et al.} developed the Mpipi model, a residue-level CG model parameterized through a combination of atomistic simulations and bioinformatics data \cite{Joseph2021}. By focusing on pi-pi and cation-pi interactions, they successfully reproduced the experimental phase behavior of a set of IDPs and of a poly-arginine/poly-lysine/RNA system. However, due to the absence of explicit protein-solvent interactions, it leads to a poor representation of protein solubility upon temperature modifications. Tesei \textit{et al.} developed CALVADOS, a CG model trained using Bayesian learning of experimental data, including SAXS and Paramagnetic Resonance Enhancement (PRE) NMR data \cite{Tesei2021_2}. CALVADOS is able to capture the phase separation of the low complexity domains of FUS, Ddx4, hnRNPA1 and LAF proteins, and has been used to calculate the ensembles of the IDRs within the human proteome \cite{Tesei2024}.

Finally, CG models with a higher resolution, such as the Martini model, have been also used to study monomeric IDPs/IDRs and their ensembles \cite{Marrink2023}. Although originally developed to model folded proteins, the model accurately reproduces the phase separation of disordered regions, the partitioning of molecules and the simulation of chemical reactions within the ensembles \cite{Tsanai2021, Brasnett2024}. Thomasen \textit{et al.} demonstrated that the Martini CG model tends to underestimate the global dimensions of disordered proteins \cite{Thomasen2022}; by augmenting the protein-water interactions, they successfully reproduced SAXS and PRE data for a set of disordered and multidomain proteins. Multiscale approaches that consider a combination of AA/CG resolutions have also been proposed.\cite{Ribeiro_Filho2022,Ingolfsson2023}

\subsection{Machine-learning Methods}
With the advent of AlphaFold2 \cite{Jumper2021}, RoseTTAFold \cite{Baek2021}, and ESMFold \cite{Lin2023}, it is clear that different ML models can produce high quality predicted structures of globular proteins. These methods use multiple sequence alignments (MSA) and clustering to generate different conformers of folded proteins, but caution is warranted for IDPs for which MSAs are not fully applicable.\cite{Wayment_Steele_2023, Sala_2023} In a recent study, Alderson and co-workers suggested that when high confident predictions are made for IDRs from AlphaFold, these conformations likely reflect those in conditionally folded states.\cite{Alderson2023} However, many IDRs from AlphaFold2 seem to be built as ribbons in a way that only satisfies the steric clash restraints rather than any secondary structure, or global shape parameters and are designated as low-confidence predictions. 

The large influx of recent ML methods to predict ensembles of disordered proteins can use pre-existing protein structure prediction algorithms or by training on CG or AA MD data. Denoising diffusion models can be used to generate coarse-grained IDP/IDR ensembles, as in the work of Taneja and Lasker \cite{Taneja2024}. Phanto-IDP, a generative model trained on MD trajectory data that uses a modified graph variational auto-encoder, can generate 50,000 conformations of a 71-residue IDP within a minute \cite{Zhu2023}. Similarly, Janson and coworkers have trained a Generative Adversarial Network (GAN) based on CG and AA simulations of IDPs (idpGAN) \cite{Janson2023}, in which the model learns the probability distribution of the conformations in the simulations to draw new samples based on sequences of IDPs. Lotthammer and coworkers have trained a Bidirectional Recurrent Neural Network with Long Short-Term Memory cells (BRNN-LSTM) on IDP ensembles generated with the Mpipi CG force field, to predict properties of disordered proteins (ALBATROSS).\cite{Lotthammer2024} There have also been methods to generate ensembles of protein structures using AI-augmented MD simulations with the option to use AlphaFold in the post-processing stage in a method called "Re-weighted Autoencoded Variational Bayes for Enhanced Sampling" (AlphaFold-RAVE) \cite{Vani2023}.

\section{Ensemble Validation}
Although there are many creative solutions to generate conformer ensembles of IDPs, the ensembles must be validated against experimental data to further improve upon their utility. Extracting structural and dynamic information from IDPs and IDRs can be done using a variety of solution-based experimental procedures, although these observables tend to be highly averaged. Hence computational tools go hand-in-hand with experiments, providing structural detail while being validated against back-calculation for each experimental data type. Additionally, ensemble generation pipelines can make use of back-calculated data to enforce experimental restraints during an integrative modeling process.\cite{Bonomi2017}

\subsection{Experimental Observables}
NMR spectroscopy is a powerful technique used to provide structural insights and dynamic properties of disordered proteins at close-to physiological conditions \cite{Bhowmick2016, Karamanos2022}. A key NMR observable is the chemical shift that provides structural information that is sensitive to functional groups and their environment. Hence methods for predicting secondary structure propensities from chemical shifts have been developed, such as SSP \cite{Marsh2006}, $\delta$2D\cite{Camilloni2012}, and CheSPI\cite{Nielsen2021} used to inform local fractional secondary structure of an ensemble. Back-calculators that predict atomic chemical shift values of protein structures include ML feature-based approaches such as SPARTA+ \cite{Shen2010}, ShiftX2 \cite{Han2011}, and UCBShift \cite{Li2020}.

J-couplings provide essential information about the connectivity and bonding between nuclei, with 3-bond J-couplings, $^{3}J$, reporting on torsion angle distributions, and can reveal information about the timescales of molecular motions in IDPs \cite{Kosol2013}. In the case of PTM-stabilized or folding-upon-binding category of IDPs, changes in $^{3}J$-coupling patterns may indicate the presence of secondary structure elements or structural motifs. The back-calculation of J-coupling data is relatively straight forward by using the Karplus equation \cite{Karplus1963} and the protein backbone torsion angles. An application of the Karplus equation to back-calculate $^{3}J$-coupling constants can be found in the `jc` module in SPyCi-PDB \cite{LiuJOSS2023} as well as in optimization algorithms such as X-EISD \cite{Lincoff2020} described below.

PRE and nuclear Overhauser effect (NOE) experiments provide distance information between pairs of residues, such as information about transient interactions and conformational changes \cite{Johnson2021}, and the presence of persistent local structures or long-ranged contacts that stabilize IDPs/IDRs and protein complexes and condensates\cite{Anglister2016}. Due to the conformational heterogeneity of IDPs/IDRs, both measured PRE and NOE data are averaged values across the conformational landscape. Furthermore, there are different interpretations of how to use PREs and NOEs in the form of distance restraints or as dynamical observables, which is a consideration for the back-calculation approach. When PRE and NOE values are interpreted as inter-atomic distances, the back-calculation is straightforward using Euclidean distance formulas as seen in SPyCi-PDB\cite{LiuJOSS2023}. When interpreted as dynamic quantities, the NOE back-calculation is a time correlation function that can more accurately represent the experimental observable as shown by Ball and co-workers. \cite{Ball2014} Ideally the back-calculation of PREs would be in the form of intensity ratios or rates, as they are native to the experimental protocol and thus subject to less error due to different interpretations of the conversion from $T_{1}$/$T_{2}$ relaxation rates to distances using the Solomon-Bloembergen equation \cite{Solomon1955}. An example of back-calculating PRE ratios can be seen in DEER-PREdict \cite{Tesei2021}, where intensity ratios are estimated instead of distances.

NMR relaxation experiments can describe the dynamics of disordered proteins. The transverse relaxation time ($T_{2}$) and the associated $R_{2}$ relaxation rates (i.e. $R_{2}$ = 1/$T_{2}$) can be obtained by from NMR experiments in which shorter $T_{2}$ values are associated with increasing protein dynamics. The $S^{2}$ order parameter provides a measure of the amplitude of motion on a fast picosecond-nanosecond timescale which ranges from 0 (completely disordered) to 1 (folded). $S^{2}$ values can be derived from relaxation data ($T_{1}$, $T_{2}$, NOE measurements) and is useful for identifying regions of a protein that are more flexible/disordered. As seen in the work from Naullage \textit{et al.}, $R_{2}$ and $S^{2}$ values were used in addition to other experimental restraints to elucidate a structural ensemble for the unfolded state of the drkN SH3 domain \cite{Naullage2022}. Backbone $^{15}N$ $R_{2}$ values in ENSEMBLE\cite{ENSEMBLEMarsh2011} are calculated as the number of heavy atoms within 8 Å. The $R_{2}$ restraint has also been used within the work of Marsh \textit{et al.}, where the authors model three nonhomologous IDPs, I-2, spinophilin, and DARPP-32.\cite{Marsh2010}

Double Electron-Electron Resonance (DEER), also known as pulsed electron-electron double resonance (ELDOR), is an EPR spectroscopy that uses site-directed spin labeling \cite{LeBreton2015} to explore flexible regions of proteins to measure distances in the range of 18-60 Å \cite{Jeschke2012} between two spin-labeled sites. Due to the effective distance measurements, DEER can help characterize the range of distances sampled by different regions within an IDP. Furthermore, DEER can be used to study conformational changes that occur when IDPs undergo folding upon binding to their interaction partners and provide insights into the structural transition from a disordered to ordered state \cite{LeBreton2015, Evans2023}. Using a rotamer library approach, the Python software package called DEER-PREdict effectively back-calculates electron-electron distance distributions from conformational ensembles \cite{Tesei2021}. Additionally, a plugin for the PyMOL molecular graphics system \cite{PyMOL} can estimate distances between spin labels on proteins in an easy-to-use graphical user interface format \cite{Hagelueken2012}.

Single-molecule Förster resonance energy transfer (smFRET) is a popular fluorescence technique used to study the conformational dynamics of disordered protein systems.\cite{Gomes2020}. It is commonly referred to as a "spectroscopic ruler" with reported uncertainties of the FRET efficiency $\le$ 0.06, corresponding to an interdye distance precision of $\le$ 2 Å and accuracy of $\le$ 5 Å \cite{Agam2023}. Because of high uncertainties for distances between inter-residue donor and acceptor fluorophores, it is better to back-calculate the FRET efficiencies $\langle$E$\rangle$ instead. Prediction of FRET efficiencies can be done using a recently developed Python package FRETpredict \cite{Montepietra2023}, for which the same rotamer library approach used to predict DEER and PRE values (using DEER-PREdict) is used here to obtain either a static, dynamic, or average FRET efficiency of an conformer ensemble. Calculations of FRET efficiencies can also be done through MD simulations\cite{Shaw2020}; AvTraj is an open-source program to post-predict FRET efficiencies from MD trajectories \cite{Dimura2016} of conformer ensemble models.

Photoinduced electron transfer (PET) coupled with fluoresence correlation spectroscopy (FCS), known as PET-FCS, identifies the contact ($\le$ 10 Å) formation rate between a fluorophore and a quencher (an aromatic residue or another dye) \cite{Abyzov2022}. Furthermore, FCS measurements are not the efficiencies of energy transfer as reported for FRET but FCS measures the amplitude of the intensity of fluorescence over time as a reporter for protein diffusion and concentration. \cite{Cubuk2022} Due to the time dependency of PET-FCS however, back-calculation techniques are linked to MD simulation data. For example, by exploiting the ns-$\mu$s timescale of protein chain dynamics, PET-FCS is a valuable experimental protocol to study disordered protein kinetics as presented in the example of studying the disordered N-terminal TAD domain of the tumor suppressor protein p53 \cite{Lum2012}. 

Finally, for the measurement of global structural dimensions, small angle X-ray scattering (SAXS) can determine the radius of gyration, $R_{g}$, while the hydrodynamic radius ($R_{h}$) can be measured by using NMR pulsed field gradient diffusion or by Size Exclusion Chromatography (SEC). SAXS is commonly reported for isolated IDPs or IDPs/IDRS in complexes under a variety of \textit{in vitro} experimental conditions \cite{DaVela2020}. For disordered protein systems, SAXS scattering data can also unveil disorder-to-order transitions, and assist in ensemble modeling by restraining global dimensions \cite{Kikhney2015}. The value of the obtained $R_{h}$ value reflects the protein's size in a solvent, accounting for its shape and the surrounding medium's viscosity. SAXS measurements can also be used to derive the hydrodynamic radius ($R_{h}$). 

Although back-calculations can be performed for $R_{g}$ with relative ease from structural ensembles, predicting SAXS intensity profiles requires increased biophysical rigor as presented in CRYSOL \cite{ManalastasCantos2021}, AquaSAXS \cite{Poitevin2011}, Fast-SAXS-pro \cite{Ravikumar2013}, and FoXS \cite{SchneidmanDuhovny2013}.Back-calculations of $R_{h}$ values, however, have been more ambiguous due to the various solvent effects that change the distribution of $R_{h}$ values for a given protein. Popular back-calculation methods for $R_{h}$ include: i) HYDROPRO \cite{GarcadelaTorre2000}, by implementing a bead modeling algorithm that expands the volume of atomic spheres based on their covalent radii, ii) HullRad \cite{Fleming2018}, which uses the convex hull method to predict hydrodynamic properties of proteins, and iii) Kirkwood-Riseman equation \cite{Kirkwood1954} by using the center of mass of each residues instead of the C$\alpha$.

\subsection{Statistical subsetting/reweighting methods}
Given an "initial" IDP/IDR ensemble generated using one of the different ensemble generation protocols, these must be modified by imposing that the IDP/IDR structural ensembles agree with available experimental data.\cite{Czaplewski2021} This integrative biology step has often relied on reweighting methods such as maximum-parsimony (MaxPar) or maximum-entropy (MaxEnt) approaches to create better validated IDP/IDR structural ensembles. MaxEnt is a probabilistic method based on the principle of maximizing the degree of disorder while also satisfying a set of experimental restraints, giving conformations higher weights when they are more consistent with an experimental observable. MaxPar approaches instead focus on finding the simplest IDP/IDR ensemble by minimizing the number of conformations in the ensemble to those that satisfy the experimental constraints. Usually, MaxPar approaches are used when there is more confidence in the available experimental data, where the goal is to simplify the ensemble to a minimum set of conformations \cite{Bonomi2017, GamaLimaCosta2022}. For a more in-depth comparison of MaxEnt and MaxPar approaches of ensemble reweighting, please refer to the review of Bonomi et al. \cite{Bonomi2017}. 

Examples of software packages and methods that fall into the MaxPar umbrella include: the 'ensemble optimization method' (EOM) \cite{EOMBernad2007}, the 'selection tool for ensemble representations of intrinsically disordered states' (ASTEROIDS) \cite{ASTEROIDSNodet2009}, and the 'sparse ensemble selection' (SES) method \cite{SESBerlin2013}. MaxEnt approaches for ensemble reweighting include the 'ensemble-refinement of SAXS' (EROS) \cite{EROSRycki2011}, 'convex optimization for ensemble reweighting' (COPER) \cite{COPERLeung2015}, and ENSEMBLE \cite{ENSEMBLEKrzeminski2012}. But more recent developments of the MaxEnt approach have opted to include Bayesian inference using back-calculated and experimental data. Since this statistical framework allows for combining different sources of data, it allows for independent accounting for experimental and back-calculator errors. The use of Bayesian statistics can be found in algorithms that re-weight ensembles directly from MD simulations such as the 'Bayesian energy landscape tilting' (BELT) method \cite{Beauchamp2014}, 'Bayesian Maximum Entropy' (BME) \cite{BMEBottaro2020}, and 'Bayesian inference of conformational populations' (BICePs) \cite{BICePsRaddi2023}. The Bayesian Extended Experimental Inferential Structure Determination (X-EISD) for IDP ensemble selection evaluates and optimizes candidate ensembles by accounting for different sources of uncertainties, both back-calculation and experimental, for smFRET, SAXS, and many NMR experimental data types\cite{Brookes2016, XEISDLincoff2020}.

\subsection{ML Ensemble Generation and Validation}
Creating IDP ensembles that agree with experimental data such as NMR, SAXS, and smFRET has typically involved operations on static structural pools, i.e. by reweighting the different sub-populations of conformations to agree with experiment.\cite{Hummer2015,Brookes2016,Bonomi2017,Bottaro2020} But if important conformational states are absent, there is little that can be solved with subsetting and reweighting approaches for IDP/IDR generation and validation. Recently, Zhang and co-workers bypassed standard IDP ensemble reweighting approaches and instead directly evolved the conformations of the underlying structural pool to be consistent with experiment, using the generative recurrent-reinforcement ML model, termed DynamICE (Dynamic IDP Creator with Experimental Restraints).\cite{Zhang2023} DynamICE learns the probability of the next residue torsions X$_{i+1}$=[$\phi_{i+1}$,$\psi_{i+1}$, $\chi1_{i+1}$, $\chi2_{i+1}$,...] from the previous residue in the sequence X$_{i}$ to generate new IDP conformations. As importantly, DynamICE is coupled with X-EISD in a reinforcement learning step that biases the probability distributions of torsions to take advantage of experimental data types. DynamICE has used J-couplings, NOE, and PRE data to bias the generation of ensembles that better agree with experimental data for $\alpha$-synuclein, $\alpha-$beta, Hst5, and the unfolded state of SH3\cite{Zhang2023}. 

\section{Software and Data Repositories for IDPs/IDRs}
Table \ref{table:tools} presents a summary of current ensemble modeling (M), experimental validation (V), and statistical reweighting/filtering (R) approaches that are available for both users and developers. We also  highlight software that has the ability to model protein-protein interactions. Emerging developments are being made to exploit the intra- and intermolecular contact data in the RCSB PDB to model dynamic complexes of disordered proteins \textit{de novo}, and current docking methods exist for some disordered protein systems. An example of a docking method specifically designed for disordered proteins involved in complexes with folded proteins is IDP-LZerD \cite{Christoffer2020}. 

Coupled with software, curated databases play an immensely important role in IDP/IDR generation and validation. The Biological Magnetic Resonance Data Bank (BMRB) is an international open data repository for biomolecular NMR data\cite{Hoch2023}, and the small-angle scattering biological data bank (SASBDB) \cite{Kikhney2020} have seen increases in IDP/IDR data depositions. Examples of curated sequence and structural IDP ensemble coordinates include the protein ensemble database (PED) \cite{Ghafouri2023}, DisProt \cite{Quaglia2021}, the ModiDB \cite{Piovesan2022}, the database of disordered protein prediction D2P2 \cite{Oates2012}, the DescribeProt database \cite{Zhao2020}, and the Eukaryotic Linear Motif Resource (ELM) \cite{Kumar2021}.

\rowcolors{2}{gray!25}{white}
\footnotesize
\begin{longtable}{P{0.75cm}|P{0.75cm}|P{2.5cm}|P{5cm}|P{6cm}}
\hiderowcolors
\caption{\textbf{Computational tools for studying ensembles of disordered proteins.} Categorized by ensemble modeling (M), experimental validation (V), and statistical reweighting/filtering (R). *Asterisk labeled tools have the ability to model protein-protein interactions. Computational tools are sorted in chronological order from the latest published software at the time of writing (August 2024).}\\
\textbf{Year} & \textbf{Type} & \textbf{Name} & \textbf{Accessibility} & \textbf{Short Description}\\ 
\showrowcolors
\hline 2024 & M & PTM-SC & github.com/THGLab/ptm\_sc & Packing of AA side chains with PTMs.\\
\hline 2024 & V & FRETpredict & github.com/KULL-Centre/FRETpredict & Calculate FRET efficiency of ensembles and MD trajectories.\\
\hline 2024 & M & Phanto-IDP & github.com/HFChenLab/PhantoIDP & Generative ML model to reconstruct IDP ensembles.\\
\hline 2023 & M & IDPConfGen* & github.com/julie-forman-kay-lab/IDPConformerGenerator & Platform to generate AA ensembles of IDPs, IDRs, and dynamic complexes.\\
\hline 2023 & V & SPyCi-PDB & github.com/julie-forman-kay-lab/SPyCi-PDB & Platform to back-calculate different types of experimental data from IDP ensembles.\\
\hline 2023 & M & DynamICE & github.com/THGLab/DynamICE & Generative ML model to generate new IDP conformers biased towards experimental data.\\
\hline 2023 & M & idpGAN & github.com/feiglab/idpgan & ML ensemble generator for CG models of IDPs.\\
\hline 2021 & M & DIPEND & github.com/PPKE-Bioinf/DIPEND & Pipeline to generate IDP ensembles using existing software.\\
\hline 2021 & V & DEER-PREdict & github.com/KULL-Centre/DEERpredict & Calculates DEER and PRE predictions from MD ensembles.\\
\hline 2020 & M & IDP-LZerD* & github.com/kiharalab/idp\_lzerd & Models bound conformation of IDP to an ordered protein.\\
\hline 2020 & R & X-EISD & github.com/THGLab/X-EISDv2 & Bayesian statistical reweighting method for IDP ensembles using maximum log likelihood.\\
\hline 2020 & M & FastFloppyTail & github.com/jferrie3/AbInitioVO-and-FastFloppyTail & Ensemble generation of IDPs and terminal IDRs using Rosetta foundations.\\
\hline 2020 & V & UCBShift & github.com/THGLab/CSpred & ML chemical shift predictor.\\
\hline 2020 & R & BME & github.com/KULL-Centre/BME & Bayesian statistical reweighting method for IDP ensembles using maximum entropy.\\
\hline 2018 & V & HullRad & http://52.14.70.9/ & Calculating hydrodynamic properties of a macromolecule.\\
\hline 2017 & VR & ATSAS V3 & embl-hamburg.de/biosaxs/crysol.html & Collection of software.\\
\hline 2017 & MVR & NMRbox & nmrbox.nmrhub.org/ & Collection of software.\\
\hline 2013 & MVR & ENSEMBLE & pound.med.utoronto.ca/\textasciitilde{}JFKlab/ & Modeling ensembles of IDPs using experimental data.\\
\hline 2012 & M & Flexible-meccano & ibs.fr/en/communication-outreach/scientific-output/software/flexible-meccano-en & Modeling conformer ensembles of IDPs using backbone Phi-Psi torsion angles.\\
\hline 2012 & V & AvTraj & github.com/Fluorescence-Tools/avtraj & Calculate FRET observables from MD trajectories.\\
\hline 2009 & R & ASTEROIDS & closed source- contact authors & Stochastic search of conformers that agree with experimental data.
 \label{table:tools}
\end{longtable}

\newpage

\section{Conclusion and Future Directions}
For folded proteins, crystal structures provide concrete, predictive, and yet conceptually straightforward models that can make powerful connections between structure and protein function. X-ray data has informed the deep learning model AlphaFold2 such that the folded protein structure problem is in many (but not all) cases essentially solved. IDPs and IDRs require a broader framework to achieve comparable insights, requiring significantly more than the one dominant experimental tool and computational analysis approach. Instead, a rich network of interactions must occur between the experimental data and computational simulation codes, often facilitated by Bayesian analysis and statistical methods that straddle the two approaches, e.g., by constraining the simulations and/or to provide confidence levels when comparing qualitatively different structural ensembles for the free IDP and disordered complexes and condensates. 

Given the diverse range of contexts in which disordered proteins manifest their function, the development of a comprehensive software platform capable of aiding IDP/IDR researchers holds immense promise. Crafting scientific software with a focus on best practices, modularity, and user-friendliness will not only enhance its longevity, ease of maintenance, and overall efficiency, but also contribute to its utility as a plug-and-play software pipeline for the diversity of IDP/IDR studies. An inspirational example from the NMR community is the creation of NMRBox\cite{Baskaran2022}, which provides software tools, documentation, and tutorials, as well as cloud-based virtual machines for executing the software. 

As emphasized earlier, the modeling of multiple disordered chains within a complex or condensate is the next frontier. However, many of these powerful computational tools and models have been developed for the characterization, analysis, and modeling of single chain IDPs/IDRs. More focus is needed for software that can handle disordered proteins that are part of dynamic complexes. Most often modeling dynamic complexes and condensates can be done by using CG MD simulations, and given a template, modeling of dynamic complexes could be done within the Local Disordered Region Sampling (LDRS) module within IDPConformerGenerator \cite{Teixeira2022, Liu2023}. Although AlphaFold multimer \cite{Evans2021} seems promising for multi-chain complexes of folded proteins, its core MSA protocol cannot be readily exploited for disordered proteins. The developers of HADDOCK \cite{vanZundert2016}, a biophysical and biochemical driven protein-protein docking software are working on HADDOCK v3 to dock IDP/IDR conformer ensembles to other disordered or folded templates to generate ensembles of complexes. It may also be possible to model condensates this way by expanding the docking between protein chains.

Centralized data repositories and resources for IDPs/IDRS are important for several reasons. They help establish community standards for IDP/IDR data quality, and hence control the quality of the resulting IDP ensemble generation and validation. Because ML methods require robust training datasets, there is a great need to expand on both experimental and MD trajectory data deposition for disordered protein and proteins with disordered regions. For wet-lab scientists, we would like to have a call-to-action for authors in future publications to have clear deposition of experimental data in order to develop better \textit{in silico} back-calculators and ultimately ensure better ensemble generation, validation, and reweighting protocols. For computational scientists, we would like to ask the community for a standardized method for evaluating ML/AI based tools for ensemble prediction and generation. Due to the rapid rate of silicon development, it is important to have a rigorous benchmark for new validation and ensemble generation protocols.

\section{Acknowledgements}
\noindent
We thank the National Institutes of Health (2R01GM127627-05) for support of this work. J.D.F.-K. also acknowledges support from the Natural Sciences and Engineering Research Council of Canada (NSERC, 2016-06718) and from the Canada Research Chairs Program. 

\section{Author Contributions Statement}
\noindent
Z.H.L., M.T. and T.H.G. wrote the paper. All authors discussed the perspective topics, references, and made comments and edits to the manuscript.

\section*{Competing Interests Statement}
\noindent
The authors declare no competing interests.

\bibliography{references}
\end{document}